\documentclass[twocolumn,pra,showpacs,amsmath,amssymb]{revtex4}

\usepackage{graphicx}
\usepackage{dcolumn}
\usepackage{bm}

\begin{document}

\title{Effects of photon losses on phase estimation near the Heisenberg limit\\
using coherent light and squeezed vacuum}

\author{Takafumi Ono$^1$}
\email{d075972@hiroshima-u.ac.jp}
\author{Holger F. Hofmann$^{1,2}$}%
\email{hofmann@hiroshima-u.ac.jp}
\affiliation{%
$^1$Graduate School of Advanced Sciences of Matter, Hiroshima University,\\
Kagamiyama 1-3-1, Higashi Hiroshima 739-8530, Japan 
}%
\affiliation{$^2$JST,CREST, Sanbancho 5, Chiyoda-ku, Tokyo 102-0075, Japan}


\begin{abstract}
Two path interferometry with coherent states and squeezed vacuum can achieve 
phase sensitivities close to the Heisenberg limit when the average photon number
of the squeezed vacuum is close to the average photon number of the coherent light.
Here, we investigate the phase sensitivity of such states in the presence of photon
losses. It is shown that the Cramer-Rao bound of phase sensitivity can be 
achieved experimentally by using a weak local oscillator and photon counting 
in the output. The phase sensitivity is then given by the Fisher information $F$ of the
state. In the limit of high squeezing, the ratio $(F-N)/N^2$ of 
Fisher information above shot noise to the square of the average photon number $N$
depends only on the average number of photons lost, $n_{\mathrm{loss}}$, and the
fraction of squeezed vacuum photons $\mu$. For $\mu=1/2$, the effect of losses
is given by $(F-N)/N^2=1/(1+2 n_{\mathrm{loss}})$. The possibility of increasing the
robustness against losses by lowering the squeezing fraction $\mu$ is considered and
an optimized result is derived. However, the improvements are rather small, with
a maximal improvement by a factor of two at high losses.
\end{abstract}

\pacs{42.50.St, 
03.65.Ta, 
06.20.Dk, 
42.50.Dv 
}

\maketitle

\section{Introduction}
\label{sec1}

Quantum states of light can improve the sensitivity of phase measurements
beyond the limits that apply to classical light sources.
The phase sensitivity of coherent light (and hence of all classical light)
is limited by the shot noise of independent photon detection events to
the standard quantum limit of $\delta \phi^2 = 1/N$. This limit can be
overcome by using the multi-photon coherences of non-classical light
\cite{Cave81,Yur86,Xia87,Hol93,San95,Pari95,Ber00,Wan05,Pez06,Hof06,Gio06,Ono08,Pez09}.
For two mode $N$-photon systems, the highest possible phase sensitivity is achieved
by maximally path-entangled states, which are superposition states 
$(|N;0 \rangle + | 0;N \rangle)/\sqrt{2}$ where all photons are either in one path 
or in the other path of a two path interferometer\cite{Lee02,Steu02,Eda02,Kok02,Fiu02,Pry03,Wal04,Mit04,Nag07}. The phase sensitivity of these states
defines the Heisenberg limit of $\delta \phi^2 = 1/N^2$. Since no $N$-photon
states achieve a higher phase sensitivity, this is the absolute limit of phase
estimation for a fixed photon number $N$ \cite{Gio06}.

Unfortunately, it is rather difficult to generate maximally path-entangled states using
the available sources of non-classical light \cite{Hof04,Cabl07,Niel07}. It was therefore 
a significant discovery that the interference of a coherent state and a squeezed
vacuum produces a high fraction of maximal path-entangled states when the average
photon number from the squeezed vacuum is about equal to the average photon number
of the coherent light \cite{Hof07,Pezz08}. In particular, Pezze and Smerzi showed that 
conventional two-path interferometry can achieve phase sensitivities close to the
Heisenberg limit even in the presence of fluctuating total photon number 
\cite{Pezz08}. These results seem to put Heisenberg limited phase estimation
within the reach of well-established quantum technologies. However, maximally 
path-entangled states are very sensitive to photon losses, since the loss of a
single photon can completely randomize the multi-photon coherence between the paths
\cite{Sean08,Dor09,Dem09}.
We can therefore expect that the Heisenberg-limited phase sensitivity achieved 
by coherent light and squeezed vacuum will rapidly decline as photon losses increase.

In the following, we investigate the effect of photon losses on the phase 
sensitivity of the two-mode states generated by interference of coherent
light and squeezed vacuum in detail. Assuming equal losses in both optical modes, we
derive the mixed state after losses and find an optimal phase estimator based
on the general analysis for quantum metrology using mixed states \cite{Braun94}.
We find that the Cramer-Rao bound giving the maximal phase sensitivity of
the state can be achieved by a simple experimental setup using a weak local
oscillator field and photon detection. It is therefore possible to obtain a
phase sensitivity equal to the Fisher information $F$ in an experimentally feasible
setup using only linear optics and photon detection. In the limit of high squeezing,
the ratio $(F-N)/N^2$  of Fisher information above shot noise to 
the square of the average photon number $N$ depends only on the fraction of squeezed 
vacuum photons $\mu$ and the average number of photons lost $n_{\mathrm{loss}}$. 
Thus the effect of photon losses on phase 
sensitivities close to the Heisenberg limit has the same dependence on the average number 
of photons lost, regardless of the total average photon number $N$. 
In particular, photon losses reduce the phase sensitivity for Heisenberg-limited 
estimation at $\mu=1/2$ by a factor of $1/(1+2n_{\mathrm{loss}})$.
Finally, we investigate the possibility of improving the Fisher information by
optimizing the squeezing fraction $\mu$ for a given number of photons lost. 
However, the result shows only small improvements, approaching a maximal increase
of Fisher information by a factor of two in the limit of high photon losses.

\begin{figure}[t]
\begin{picture}(400,150)
\put(-30,20){\makebox(400,150)
{\vspace*{-18cm} 
\scalebox{1}[1]{\includegraphics{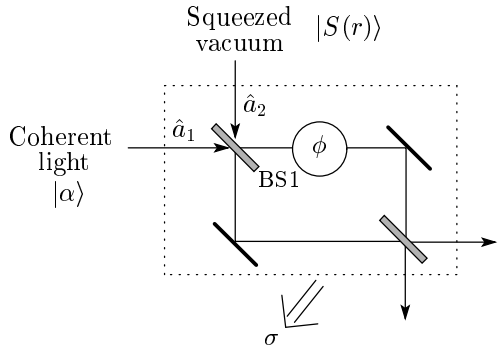}}}}
\end{picture}
\caption{\label{fig1}
Illustration of Heisenberg limited estimation of a small phase shift 
$\phi$ with coherent light and squeezed vacuum in the presence of linear
losses. The probability that any given photon is lost between generation and
detection is given by the loss fraction $\sigma$.}
\end{figure}

\section{Effects of losses on the two mode squeezed-coherent state}
\label{sec2}

Fig. \ref{fig1} shows a possible experimental setup realizing Heisenberg limited
phase estimation with coherent light and squeezed vacuum in the input ports of
a two-path interferometer.
Initially, mode $\hat{a}_1$ is in a coherent state $|\alpha \rangle$ and mode
$\hat{a}_2$ is in a squeezed vacuum state $|S(r) \rangle=\hat{S}(r) |\mbox{vac.} \rangle$,
where $\hat{S}(r) = \mathrm{exp}[ 1/2( r\hat{a}_2 \hat{a}_2 - r^* \hat{a}_2^{\dagger} 
\hat{a}_2^{\dagger} )]$ is the squeezing operator. Interference at the
initial beam splitter then results in multi-photon coherences between the two
paths inside the interferometer, as discussed in \cite{Hof07,Pezz08}. 
However, photon losses occurring at any point between the generation and the 
detection of the light fields will reduce these multi-photon coherences.
In the following, we assume linear losses with equal loss rates in the two modes.
It is then possible to represent the losses by the loss fraction
$\sigma$, defined as the probability that any given photon is lost between 
generation and detection.

The effects of linear losses on the two input modes correspond to interference with
a vacuum state, followed by a trace over the modes representing the losses. 
Since photon losses from orthogonal modes are statistically independent, 
it is possible to consider the effect of photon losses on the two input modes 
separately. 
For the coherent state, the losses simply reduce the amplitude $\alpha$ by a factor 
of $\sqrt{1-\sigma}$, so that the output amplitude is
$\alpha_{\mathrm{red.}} = \sqrt{1-\sigma}\alpha$ and 
the density matrix $\hat{\rho}_1$ of mode $\hat{a}_1$ after losses is
\begin{equation}
\hat{\rho}_{1} = |\alpha_{\mathrm{red.}} \rangle \langle \alpha_{\mathrm{red.}} |.
\end{equation}
In the case of the squeezed vacuum, losses change the variances of the quadrature
components $\hat{x}_2$ and $\hat{y}_2$ of the field mode 
$\hat{a}_2= \hat{x}_2 + i \hat{y}_2$. 
The output is a Gaussian state with quadrature variances of
\begin{eqnarray}
\label{variance1}
4 \Delta x_2^2 &=& \sigma + (1-\sigma) \mathrm{e}^{-2r}
\nonumber\\
4 \Delta y_2^2 &=& \sigma + (1-\sigma) \mathrm{e}^{2r}.
\end{eqnarray}
In general, a Gaussian mixed state defined by the variances $\Delta x_2^2$ and
$\Delta y_2^2$ can be described by a squeezed thermal state,
\begin{equation}
\label{squeezed thermal state}
\hat{\rho}_2 = \hat{S}(r_{\mathrm{red.}}) \hat{\rho}_{\mathrm{th}}(\lambda) \hat{S}^{\dagger}(r_{\mathrm{red.}}),
\end{equation}
where the thermal state is given by
\begin{equation}
\hat{\rho}_{\mathrm{th}} (\lambda)= (1-\lambda) \sum_{n}^{\infty}{\lambda^n | n \rangle \langle n |}
\end{equation}
and $r_{\mathrm{red.}}<r$ is the reduced squeezing parameter obtained from the
ratio of the variances after losses. 
In terms of the parameters $\lambda$ and $r_{\mathrm{red.}}$, the variances of 
$\hat{\rho}_2$ are
\begin{eqnarray}
\label{variance2}
4 \Delta x_2^2 &=& \frac{1+\lambda}{1-\lambda} \mathrm{e}^{-2r_{\mathrm{red.}}}
\nonumber\\
4 \Delta y_2^2 &=& \frac{1+\lambda}{1-\lambda} \mathrm{e}^{2r_{\mathrm{red.}}}.
\end{eqnarray}
Thus it is possible to determine the values of $\lambda$ and $r_{\mathrm{red.}}$ 
corresponding to an initial squeezing parameter $r$ and a loss probability $\sigma$
from eqs. (\ref{variance1}) and (\ref{variance2}). 

The complete two mode state after losses is given by the product of the states in
mode $\hat{a}_1$ and in mode $\hat{a}_2$,
\begin{equation}
\hat{\rho} = \hat{\rho}_1 \otimes \hat{\rho}_2.
\end{equation}
The phase sensitivity achieved by this state can be analyzed using the general
formalism for mixed states \cite{Braun94}. It is then possible to determine both
the quantum Cramer-Rao bound of phase estimation and the measurement procedure that
achieves this bound in the presence of photon losses.

\section{Phase estimation with mixed states}
\label{sec3}

Quantum phase estimation is performed by measuring a phase estimator $\hat{A}$. 
If the average value of $\hat{A}$ is chosen to be zero at $\phi=0$, the phase
derivative of the average gives the ratio of the average of $\hat{A}$ and 
the small phase shift that quantum estimation seeks to detect. Thus the average
phase estimate $\langle \phi_{\mathrm{est.}} \rangle = \langle \hat{A} 
\rangle/(\partial \langle \hat{A} \rangle/\partial \phi )$ converges on the 
correct value $\phi$ as the number of measurements increases. However, each
individual measurement has a statistical error of $\delta \phi^2$ that determines
how quickly the average of the measurement results converges on the correct
value of $\phi$. In terms of this measurement error, the phase sensitivity obtained
with a specific estimator $\hat{A}$ is given by
\begin{equation}
\label{phase sensitivity}
\frac{1}{\delta \phi^2} = \frac{\left| \partial \left( \mathrm{Tr}\left\{ \hat{A} \hat{\rho} \right\} \right)/\partial \phi \right|^2}{\mathrm{Tr}\left\{ \hat{A}^2 \hat{\rho}\right\}}.
\end{equation}
An optimal phase estimator $\hat{A}$ maximizes this phase sensitivity and achieves the
quantum Cramer-Rao bound of the state $\hat{\rho}$. As was shown in \cite{Braun94}, the 
optimal estimator is given by the symmetric logarithmic derivative $\hat{G}$ of the
density matrix $\hat{\rho}$, as defined by the operator relation
\begin{equation}
\label{phase derivative1}
\frac{\partial}{\partial \phi} \hat{\rho} = \frac{1}{2} \left(\hat{\rho} \hat{G} + \hat{G} \hat{\rho} \right).
\end{equation} 
Note that this relation does not uniquely define $\hat{G}$ if $\hat{\rho}$ has eigenvalues of
zero. In that case, any operator $\hat{G}$ fulfilling eq. (\ref{phase derivative1}) is an optimal
estimator. 
The maximal phase sensitivity achieved by an optimal estimator $\hat{G}$ is equal to 
the Fisher information $F=1/\delta \phi_{\mathrm{opt.}}^2$ of the quantum state $\hat{\rho}$.
The Fisher information can be evaluated from eq. (\ref{phase sensitivity}) by using 
$\hat{A}=\hat{G}$ and eq. (\ref{phase derivative1}). The result is equal to the variance
of the optimal phase estimator $\hat{G}$,
\begin{equation}
\label{optimal phase sensitivity}
F = \mathrm{Tr} \left\{ \hat{G}^2 \hat{\rho} \right\}.
\end{equation}

Eqs. (\ref{phase derivative1}) and (\ref{optimal phase sensitivity}) summarize the results
for quantum metrology with mixed states obtained in \cite{Braun94} without the
explicit expansion into eigenstates of the density matrix used in the original derivation.
In general, these results apply to any parameter $\phi$ that changes the quantum state 
$\hat{\rho}$. In the specific case of a phase shift, $\phi$ is the parameter of a 
unitary transformation
 $\mathrm{exp}[ -i\phi \hat{h} ]$ generated by an operator $\hat{h}$. The phase derivative of
the density matrix is therefore given by the commutation relation of $\hat{\rho}$ and
$\hat{h}$,
\begin{equation}
\label{phase derivative2}
\frac{\partial}{\partial \phi} \hat{\rho} = -i \left( \hat{h} \hat{\rho} - \hat{\rho} \hat{h} \right). 
\end{equation}
To find an optimal estimator $\hat{G}$ for a given generator $\hat{h}$ and a given quantum state
$\hat{\rho}$, we have to solve eq. (\ref{phase derivative1}) using the phase derivative given
by eq. (\ref{phase derivative2}). This relation can be summarized by
\begin{equation} 
\label{estimator relation}
\frac{1}{2} \left( \hat{G} \hat{\rho} + \hat{\rho} \hat{G} \right) = -i\left( \hat{h} \hat{\rho} - \hat{\rho} \hat{h} \right).
\end{equation}
For unitary transforms, an optimal estimator $\hat{G}$ is therefore obtained when 
the anti-commutation of $\hat{G}$ and $\hat{\rho}$ has the same form as the commutation
of $\hat{h}$ and $\hat{\rho}$.

\section{Derivation of an optimal phase estimator}
\label{sec4}
We can now derive an optimal estimator for phase estimation with coherent 
light and squeezed vacuum in the presence of losses. The density matrix
was derived in sec. \ref{sec2}, and the generator $\hat{h}$ for a
two path interferometer is given by half the photon number difference 
between the two paths. In the present context, the two-mode density matrix is
a product of the two input mode density matrices. It is therefore
convenient to express the generator $\hat{h}$ in terms of the input
modes $\hat{a}_1$ and $\hat{a}_2$, which are equal superpositions of the
modes describing the two paths inside the interferometer. The photon number
difference between the two paths is then equal to an interference term
of the input modes $\hat{a}_1$ and $\hat{a}_2$. Specifically, it can be written 
as
\begin{equation}
\hat{h} = -i\frac{1}{2}\left( \hat{a}_1^{\dagger} \hat{a}_2 - \hat{a}_2^{\dagger} \hat{a}_1 \right).
\end{equation}
With this generator, eq. (\ref{estimator relation}) provides a relation between the
optimal estimator $\hat{G}$ and the two mode state $\hat{\rho}$ in terms of the creation and
annihilation operators of the input states,
\begin{equation}
\label{estimator relation2}
\hat{G} \hat{\rho} + \hat{\rho} \hat{G} = \hat{\rho} \left( \hat{a}_1^{\dagger} \hat{a}_2 - \hat{a}_2^{\dagger} \hat{a}_1 \right) - \left( \hat{a}_1^{\dagger} \hat{a}_2 - \hat{a}_2^{\dagger} \hat{a}_1 \right) \hat{\rho}.
\end{equation}
Since the density matrix can be written as a product of states in mode $\hat{a}_1$ and
mode $\hat{a}_2$, the effects of the operators $\hat{a}_1$ and $\hat{a}_2$ on the states 
$\hat{\rho}_1$ and $\hat{\rho}_2$ can be determined separately. 
It is then possible to derive a particularly simple form of $\hat{G}$ by only
considering the relations between single mode Gaussian states and the creation and annihilation operators of their respective modes.

First, we consider the effects of the annihilation and creation operators of mode $\hat{a}_1$
on the coherent state density matrix $\hat{\rho}_1$. The coherent state 
$\mid \alpha_{\mathrm{red.}}\rangle$ is a right eigenstate of the annihilation operator 
$\hat{a}_1$. It is therefore possible to replace the operator $\hat{a}_1$ 
operating from the left on $\hat{\rho}$ and the operator $\hat{a}_1^\dagger$ operating from the
right with the complex number $\alpha_{\mathrm{red.}}$. The application of
$\hat{a}_1^\dagger$ to the coherent state $\mid \alpha_{\mathrm{red.}}\rangle$
changes that state into a superposition of the original state with an amplitude of 
$\alpha_{\mathrm{red.}}$ and an orthogonal state that can be represented by a 
displaced one photon state. The creation operator can thus be written as a sum of
the coherent amplitude $\alpha_{\mathrm{red.}}$ and an operator that changes 
the initial state into an orthogonal state,
\begin{equation}
\hat{a}_1^{\dagger} = \alpha_{\mathrm{red.}} + (\hat{a}_1^{\dagger} - \alpha_{\mathrm{red.}}).
\end{equation}
It is possible to separate the relation for the estimator $\hat{G}$ given by 
eq. (\ref{estimator relation2}) into two parts, one that leaves the coherent state
unchanged, and one that describes the transition matrix elements between the coherent
state and the displaced one photon state. The separation is achieved by writing the
optimal estimator as $\hat{G} = \hat{g}_1 + \hat{g}_2$, where $\hat{g}_1$ is the 
component of the estimator associated with the transition matrix elements in mode
$\hat{a}_1$ and $\hat{g}_2$ is the component of the estimator that commutes with the
coherent state $\hat{\rho}_1$. The two relations defining $\hat{g}_1$ and $\hat{g}_2$
then read
\begin{eqnarray}
\label{estimator relation3}
\lefteqn{
\hat{g}_1 \hat{\rho} + \hat{\rho} \hat{g}_1 =}
\nonumber \\ &&
 -\left( (\hat{a}_1^{\dagger} - \alpha_{\mathrm{red.}})\hat{a}_2 \hat{\rho} - \hat{\rho} \hat{a}_2^{\dagger}(\hat{a}_1 - \alpha_{\mathrm{red.}}) \right) \\[0.5cm]
\label{estimator relation4}
\lefteqn{\hat{g}_2 \hat{\rho} + \hat{\rho} \hat{g}_2 =}
\nonumber \\ &&
 -\alpha_{\mathrm{red.}} \left( (\hat{a}_2 - \hat{a}_2^{\dagger}) \hat{\rho} - \hat{\rho}(\hat{a}_2 - \hat{a}_2^{\dagger}) \right).
\end{eqnarray}

To find $\hat{g}_1$, we make use of the fact that $(\hat{a}_1 - \alpha_{\mathrm{red.}})
\hat{\rho}=0$.
It is therefore possible to add or subtract multiples of this operator and its self-adjoint
operator to the right side of eq. (\ref{estimator relation3}) without changing the relation.
The solution for the self-adjoint operator $\hat{g}_1$ obtained in this manner is
\begin{equation}
\label{G1}
\hat{g}_1 = - \left( (\hat{a}_1^{\dagger} - \alpha_{\mathrm{red.}})\hat{a}_2 + \hat{a}_2^{\dagger} (\hat{a}_1 - \alpha_{\mathrm{red.}}) \right).
\end{equation}

To find $\hat{g}_2$, we make use of the fact that eq.(\ref{estimator relation4}) only includes
operators acting on the state in mode $\hat{a}_2$. Therefore, we only need to consider the
density matrix $\hat{\rho}_2$ of the squeezed thermal state. 
Since $(\hat{a}_2 - \hat{a}_2^{\dagger})=2 i \hat{y}_2$ corresponds to 
the anti-squeezed quadrature component, the right side of eq. (\ref{estimator relation4}) 
corresponds to the commutation relation between the quadrature $\hat{y}_2$ and the density 
matrix $\hat{\rho}_2$.
For Gaussian states, this kind of commutation relation can be converted into an anti-commutation
relation for a different quadrature component. In the case of the squeezed thermal states
given by $\hat{\rho}_2$, the conversion is given by
\begin{equation}
\label{quadrature relation}
-i\left(\hat{y}_2 \hat{\rho}_{2} - \hat{\rho}_{2} \hat{y}_2\right) = \frac{1-\lambda}{1+\lambda}  \mathrm{e}^{2r_{\mathrm{red.}}} \left(\hat{x}_2 \hat{\rho}_2 + \hat{\rho}_{2} \hat{x}_2 \right).
\end{equation}
Comparison with eq. (\ref{estimator relation4}) indicates that  $\hat{g}_2$ is a multiple 
of the quadrature component $\hat{x}_2$. Specifically, eq. (\ref{estimator relation4}) can
be solved by
\begin{equation}
\label{G2}
\hat{g}_2 = \frac{\alpha_{\mathrm{red.}}}{2 \Delta x_2^2} \; \hat{x}_2,
\end{equation}
where the coefficients $\lambda$ and $r_{\mathrm{red.}}$ have been expressed in terms of
the quadrature variance $\Delta x_2^2$ using eq. (\ref{variance2}).

The optimal estimator $\hat{G}$ is given by the sum of $\hat{g}_1$ and $\hat{g}_2$. 
Since $\hat{g}_1$ is a quadratic function of the creation and annihilation operators,
and $\hat{g}_2$ is a linear function of the operators of mode $\hat{a}_2$, it is 
possible to express $\hat{G}$ as a quadratic function of the field operators. Specifically,
the result can be written as
an interference term of $\hat{a}_2$ and a field $\hat{b}$,
\begin{equation}
\label{optimal estimator2}
\hat{G} = \hat{b}^{\dagger} \hat{a}_2 + \hat{a}_2^{\dagger}\hat{b},
\end{equation}
where the field $\hat{b}$ is given by
\begin{equation}
\hat{b}= \alpha_{\mathrm{red.}} \left( \frac{1}{4\Delta x_2^2} + 1 \right) - \hat{a}_1.
\end{equation}
Experimentally, this estimator can be realized by subtracting a coherent amplitude of 
$\alpha_{\mathrm{red.}} (1/4\Delta x_2^2 + 1)$ from the output field in $\hat{a}_1$
using interference with a local oscillator.  
\begin{figure}[t]
\begin{picture}(400,150)
\put(-60,-30){\makebox(400,150)
{\vspace*{-11cm}
\scalebox{0.9}[0.9]{\includegraphics{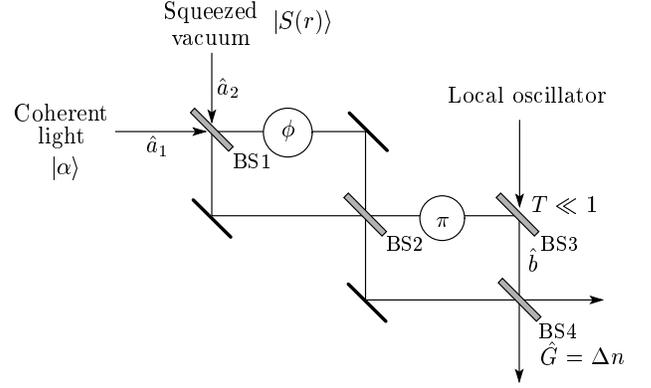}}}}
\end{picture}
\caption{\label{fig2}
Schematic setup for the observation of optimal phase sensitivity obtained by coherent light and squeezed vacuum in the presence of losses. The optimal phase estimation is realized by subtracting
a coherent amplitude of $\alpha_{\mathrm{red.}} (1/4\Delta x_2^2 + 1)$ from the output mode
$\hat{a}_1$ using a local oscillator and a beam splitter with low transmittivity $T$ (BS3).}
\end{figure}
A possible setup is shown in fig.\ref{fig2}. At zero phase shift, the output modes correspond
to the input modes. A small displacement of the coherent field mode $\hat{a}_1$ is realized by
interference with a local oscillator field at a beam splitter of very low transmittivity $T$
(BS3). Finally, the displaced mode $\hat{b}$ and the output mode $\hat{a}_2$ interfere at a 
fourth beam splitter (BS4). The phase estimator is then equal to the photon
number difference in the output. Interestingly, the estimator is a linear function of the
detected photon numbers. This is quite different from the optimal phase estimation for
pure states considered in the initial work on Heisenberg limited phase estimation with coherent 
and squeezed light, where higher order moments of the detected output photon number 
distribution were essential \cite{Pezz08}. The present setup therefore represents a 
major simplification of the phase estimation procedure for phase sensitivities close to the
Heisenberg limit, even in the pure state case where the phase sensitivity is equal to that
obtained from direct photon counting in the output.

\section{Dependence of phase sensitivity on photon losses}
\label{sec5}

As mentioned in sec. \ref{sec3}, the Fisher information of the quantum state $\hat{\rho}$ is equal 
to the expectation value of the squared estimator $\hat{G}$. Using the result of 
eq. (\ref{optimal estimator2}), the Fisher information of the squeezed-coherent state $\hat{\rho}$
is found to be
\begin{equation}
\label{optimal F}
\mbox{Tr}\{\hat{G}^2 \hat{\rho} \} = \frac{\alpha_{\mathrm{red.}}^2}{4~\Delta x_2^2} + n_2,
\end{equation}
where $n_2$ is the average number of photons in the squeezed mode $\hat{a}_2$ after losses.
Here, the effects of losses are expressed indirectly through the values of  
$n_2$, $\alpha_{\mathrm{red.}}$ and $\Delta x_2$. The specific effects of a loss
probability of $\sigma$ on the input state are given by 
$\alpha_{\mathrm{red.}} = \sqrt{1-\sigma}\alpha$, 
$n_2 = (1-\sigma) \sinh^2 r$ and eq. (\ref{variance1}). The Fisher information can then
be expressed in terms of the input amplitude $\alpha$, the input squeezing $r$, and
the loss probability $\sigma$. The result reads
\begin{equation}
\label{optimal phase sensitivity2}
F = (1-\sigma) \left(\frac{\alpha^2}{\sigma + (1-\sigma)\mathrm{e}^{-2r}} + \sinh^2 r \right).
\end{equation}
In this representation of the Fisher information, the most significant effect of the 
losses is the limitation of squeezing effects represented by $\mathrm{e}^{-2r}$.
However, it is difficult to see how this limitation relates to the maximal phase
sensitivities achieved at equal intensities of coherent light and squeezed vacuum.
It is therefore convenient to express the result in terms of average photon numbers instead.

\begin{figure}[b]
\begin{picture}(400,70)
\put(-0,-100){\makebox(400,70)
{\vspace*{-12cm}
\scalebox{1}[1]{\includegraphics{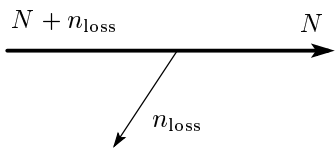}}}}
\end{picture}
\caption{\label{fig3}
Illustration of the definition of $N$ and $n_{\mathrm{loss}}$. $N$ is the
average photon number after losses.}
\end{figure}

The total average photon number after losses is given by 
$N= (1-\sigma) (\alpha^2 + \sinh^2 r)$. To evaluate the distribution of photons 
between the coherent light and
the squeezed vacuum, we introduce the squeezing fraction $\mu = (1-\sigma) \sinh^2 r/N$,
defined as the fraction of photons in the squeezed mode $\hat{a}_2$. Finally, the
effects of losses can be given in terms of the average number of photons lost,
$n_{\mathrm{loss}} = N \sigma/(1- \sigma)$. Since $N$ is the average photon number 
after losses, the total photon number of the input state is given by the sum of
$N$ and $n_{\mathrm{loss}}$, as shown in fig. \ref{fig3}. The Fisher information is 
then given by
\begin{equation}
\label{Fisher information}
F = N^2 \frac{4(1-\mu)\mu}{1 - \mathrm{e}^{-2r} + 4 \mu~n_{\mathrm{loss}}} + N.
\end{equation}
Note that this phase sensitivity can be greater than $N^2$, since the
actual Heisenberg limit for fluctuating photon numbers is given by the average of the 
squared photon number, not the square of the average photon number \cite{Hof09}. 

Since we are mainly interested in Heisenberg limited phase sensitivities with
large photon numbers, it is reasonable to assume that the squeezing levels
will be high enough to satisfy $\mathrm{e}^{-2r} \ll 1$. We can then neglect the
$r$-dependent term in eq. (\ref{Fisher information}) to obtain a particularly
simple relation between phase sensitivity and photon losses. Specifically,
the Fisher information above the standard quantum limit, $F-N$, is given by
a fraction of $N^2$ determined only by the squeezing fraction $\mu$ and the
average number of photons lost. Since this fraction does not depend on $N$, 
it provides a photon number independent expression of the effects of losses on the
phase sensitivity of squeezed-coherent states. In the following, we will refer to
this expression as the enhancement of sensitivity,
\begin{equation}
\label{Fisher dependence}
\frac{F-N}{N^2} = \frac{4\mu (1-\mu)}{1+4\mu~n_{\mathrm{loss}}}.
\end{equation}
As reported in \cite{Hof07,Pezz08}, equal intensities of squeezed vacuum and
coherent light ($\mu=1/2$) result in maximal multi-photon coherences, including a
significant fraction of maximally path-entangled states. In the absence of losses,
the enhancement of sensitivity for these pure states is $(F-N)/N^2=1$, the maximal
value that can be achieved in the limit of high squeezing. However, 
eq. (\ref{Fisher dependence}) also shows that photon losses rapidly reduce this 
enhancement of phase sensitivity. Specifically,
\begin{equation}
\left. \frac{F-N}{N^2} \right|_{\mu=\frac{1}{2}} = \frac{1}{1+2n_{\mathrm{loss}}}.
\end{equation}
This dependence of phase sensitivity on the average number of photons lost reflects 
the fact that the loss of a single photon completely 
randomizes the $N$-photon coherence of a maximally path entangled state,
irrespective of the total photon number $N$. Thus, the average loss of just half a 
photon already reduces the enhancement of sensitivity to 
half of its original value. 

\begin{figure}[t]
\begin{center}
\includegraphics[scale=0.9]{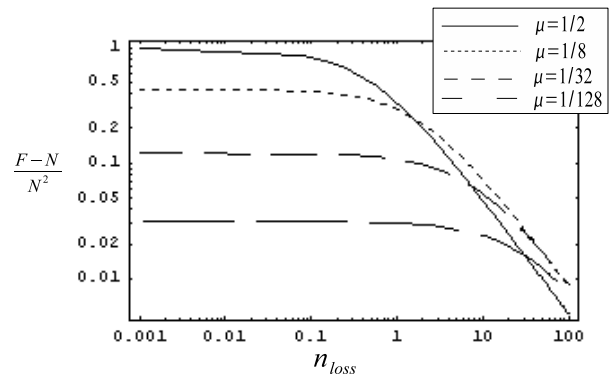}
\caption{\label{fig4}
Effects of photon losses $n_{\mathrm{loss}}$ on the enhancement of sensitivity $(F-N)/N^2$ for 
phase estimation at squeezing fractions of 
$\mu = 1/2$ and $\mu=1/8, 1/32, 1/128$.
Since photon losses reduce the enhancement of sensitivity by a factor of 
$1+4\mu~n_{\mathrm{loss}}$, states with lower squeezing fractions are more robust 
against photon losses than the states with maximal multi-photon coherence at $\mu=1/2$.
}
\end{center}
\end{figure}
Eq. (\ref{Fisher dependence}) indicates that the effect of photon losses on the 
enhancement of sensitivity $(F-N)/N^2$ decreases when the squeezing fraction $\mu$
is lowered. Specifically, photon losses reduce the enhancement of sensitivity 
by a factor of $1+4\mu~n_{\mathrm{loss}}$, defined by the product of 
squeezing fraction and photon losses. Therefore, states with lower squeezing 
fraction $\mu$ are more robust against photon losses. Fig. \ref{fig4} shows 
a comparison of the loss-dependent enhancements of sensitivity $(F-N)/N^2$ for 
different squeezing fractions $\mu \leq 1/2$. At low losses, the enhancement of
sensitivity is maximal for $\mu=1/2$ and decreases with decreasing $\mu$. As 
losses increase, the enhancement of sensitivity for $\mu=1/2$ drops to values 
below the corresponding enhancements at lower $\mu$, 
indicating that states with lower squeezing fraction can have higher 
Fisher information in the presence of losses. 
\begin{figure}[t]
\begin{center}
\includegraphics[scale=0.9]{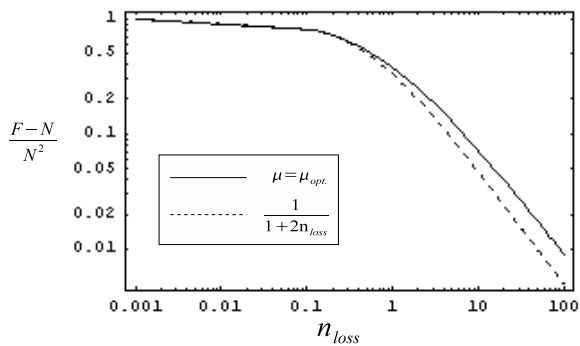}
\caption{\label{fig5}
Comparison of enhancement of sensitivity $(F-N)/N$ obtained at the 
optimized squeezing fraction $\mu=\mu_{\mathrm{opt.}}$ with the
enhancement of sensitivity obtained at $\mu=1/2$.}
\end{center}
\end{figure}
If the average number of photons lost is fixed, the highest enhancement of sensitivity given by eq. (\ref{Fisher dependence}) is found at a squeezing fraction of
\begin{equation}
\label{mopt}
\mu_{\mathrm{opt.}} = \frac{1}{4n_{\mathrm{loss}}} 
\left( \sqrt{1+4n_{\mathrm{loss}}}-1 \right).
\end{equation}
The enhancement of sensitivity at this optimal squeezing fraction 
$\mu_{\mathrm{opt.}}$ is given by
\begin{equation}
\left. \frac{F-N}{N^2} \right|_{\mu=\mu_{\mathrm{opt.}}} = \left( \frac{\sqrt{1+4n_{\mathrm{loss}}}-1}{2n_{\mathrm{loss}}} \right)^2.
\end{equation}
Fig. \ref{fig5} shows a comparison of the enhancement of sensitivity at 
$\mu_{\mathrm{opt.}}$ with the enhancement of sensitivity at $\mu=1/2$. Although 
the reduction of squeezing fraction $\mu$ results in higher enhancements of phase
sensitivity, the relative improvements seem to be rather small.
\begin{figure}[th]
\begin{center}
\includegraphics[scale=0.9]{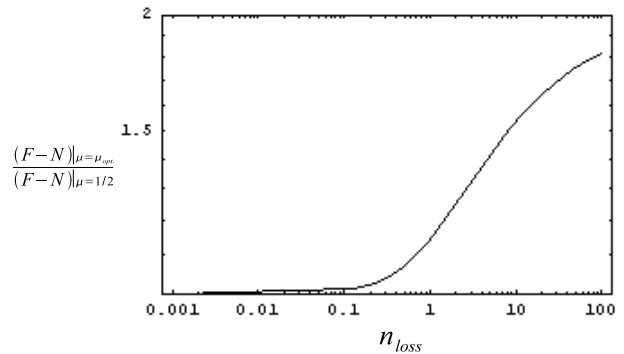}
\caption{\label{fig6}
Ratio of the sensitivity enhancement obtained at optimal squeezing fraction and
the sensitivity enhancement obtained at $\mu=1/2$. The ratio is close to one at
low losses and approaches a maximal value of two at high losses.}
\end{center}
\end{figure}
Fig. \ref{fig6} shows the improvement factor given by the ratio of the enhancement 
of sensitivity at $\mu_{\mathrm{opt.}}$ and the enhancement of sensitivity at 
$\mu=1/2$. The improvement factor is negligibly small at low losses, with a
value of only $1.072$ at average losses of half a photon. Thus, the optimization of
the squeezing fraction can do little to compensate the reduction of the enhancement
of phase sensitivity to half its value at $n_{\mathrm{loss}}=1/2$. As losses increase,
the improvement achieved by an optimization of the squeezing fraction becomes more
significant. However, the improvement is limited by its asymptotic value of 
\begin{equation}
\mathrm{lim}_{n_\mathrm{loss} \to \infty} \frac{(F-N)|_{\mu = \mu_{\mathrm{opt.}}}}{(F-N)|_{\mu=1/2}} = 2,
\end{equation}
so that the optimization of the squeezing fraction can at most double the enhancement
of phase sensitivity achieved at a squeezing fraction of $\mu=1/2$. The 
phase sensitivity achieved at $\mu=1/2$ therefore remains close to the maximal phase 
sensitivity that can be achieved with any squeezed-coherent state, even in the presence 
of very high photon losses. 
 
\section{Conclusions}

We have shown that photon losses reduce the phase sensitivity of the multi-photon
coherences obtained from interferences of equal intensities of squeezed vacuum 
and coherent light by a factor of $1+2 n_{\mathrm{loss}}$, where $n_{\mathrm{loss}}$
is the average number of photons lost. This result corresponds to the expectation
that a single photon loss randomizes the coherence of maximally path entangled states,
regardless of the total photon number $N$. A small improvement of the robustness 
against losses can be achieved by reducing the fraction of squeezed vacuum in the 
total photon number. However, the improvements are rather small and indicate that 
the robustness against losses of $N$-photon states depends mainly on their phase sensitivity, regardless of the type of state used. 

We have also shown that the Cramer-Rao
bound of squeezed-coherent states in the presence of losses can be achieved in 
experimentally feasible measurements using a weak local oscillator field, 
linear optics, and photon counting.
Interestingly, the use of the local oscillator simplifies the phase estimation 
procedure to an estimator linear in the detected photon numbers. It may therefore
present an interesting alternative to direct photon counting in the output, even 
if the improvement in phase sensitivity is negligibly small.

In general, our results confirm that phase estimation near the Heisenberg limit can be 
performed with squeezed-coherent light, but only if the average number of photons
lost can be kept low. Therefore, high efficiencies of photon transmission and
detection will be essential for quantum metrology close to the Heisenberg limit.

\section*{Acknowledgment}
Part of this work has been supported by the Grant-in-Aid program of the Japan
Society for the Promotion of Science, JSPS.


\end{document}